\documentclass[11pt]{article}

\usepackage[utf8]{inputenc}
\usepackage[T1]{fontenc}
\usepackage[margin=1in]{geometry}
\usepackage{booktabs}
\usepackage{longtable}
\usepackage{array}
\usepackage{tabularx}
\usepackage{graphicx}
\usepackage{float}
\usepackage{caption}
\usepackage[numbers,sort&compress]{natbib}
\usepackage{amsmath}
\usepackage{xcolor}
\usepackage{microtype}

\PassOptionsToPackage{hyphens}{url}
\usepackage[colorlinks=true,linkcolor=blue,citecolor=blue,urlcolor=blue,breaklinks=true]{hyperref}
\usepackage{xurl}
\graphicspath{{./}}

\newcolumntype{Y}{>{\raggedright\arraybackslash}X}
\newcolumntype{L}[1]{>{\raggedright\arraybackslash}p{#1}}
\newcolumntype{C}[1]{>{\centering\arraybackslash}p{#1}}
\title{\textbf{What quantum computer to buy?}}
\author{Alex Krasnok$^{1,2}$\thanks{Corresponding author: akrasnok@fiu.edu}\\[4pt]
\small $^{1}$Department of Electrical and Computer Engineering, Florida International University,\\ 
\small Miami, FL 33174, USA\\
\small $^{2}$Knight Foundation School of Computing and Information Sciences, Florida International University,\\ 
\small Miami, FL 33199, USA}
\date{}

\begin{document}
\maketitle

\begin{abstract}
The phrase ``buy a quantum computer'' hides several different procurement problems. An institution may be seeking cloud access for teaching, reserved capacity for research, a local instrument for hardware training, an optimization appliance, or a strategic installation that reshapes facilities, staffing, and budgets. Because these choices differ in purpose, operating burden, and useful lifetime, the decision should be framed as acquisition of \emph{quantum capability} rather than selection of a presumed hardware winner. This manuscript develops a practical procurement framework that distinguishes five capability layers, separates peer-reviewed results from commercial offerings, pricing anchors, and public roadmaps, and compares the main commercial platform families---superconducting circuits, trapped ions, neutral atoms, quantum annealing, and photonics---through the lens of institutional fit, access model, and refresh pressure. The main conclusion is that most institutions should begin with the smallest layer of capability that produces repeatable near-term value, builds internal expertise, and preserves strategic flexibility. Large on-premises systems are justified only when mission requirements, site readiness, staffing, governance, and upgrade paths are already clear.
\end{abstract}

\section{Introduction}

Buying quantum computing is a mission decision, not a hardware shopping exercise. A university may want a teaching platform that supports many students at low cost. A research center may want faster and more predictable access than a shared public queue can provide. A national laboratory may need local control, secure data handling, and close coupling to high-performance computing (HPC) systems. An industrial group may care about only one outcome: whether a benchmark workload gains anything from quantum resources at all. These are different purchases even when each buyer uses the same phrase, ``buy a quantum computer'' \citep{Preskill2018,Cerezo2021,Bharti2022}.

The market makes the decision harder because quantum hardware is heterogeneous, performance metrics are not directly comparable across vendors, and public roadmaps often move faster than institutional capital cycles. A qubit count does not reveal queue quality, software maturity, facilities burden, service model, or upgrade path. A promising research result does not guarantee a supported commercial product. A bold roadmap does not prove delivered capability. For that reason, procurement mistakes in quantum computing are often organizational before they are technical.

Public discussion still begins in the wrong place. It begins with qubit counts, vendor rankings, and claims about which platform will win in the long run. That framing gives weak procurement guidance because it ignores the operating model. A campus that needs broad student access should not start with a flagship local installation. A laboratory that needs secure, repeatable, low-latency hybrid workflows should not assume that a free cloud tier is enough. The first question is not which device looks strongest on paper. The first question is what layer of quantum capability the institution can use, support, and justify \citep{IBMPlans2026,BraketPricing2026}.

This manuscript therefore treats procurement as quantum capability acquisition. It makes four contributions. First, it defines five capability layers hidden inside the single phrase ``buy a quantum computer,'' ranging from preparation and cloud access to strategic local ownership. Second, it separates four evidence classes that buyers often mix together: peer-reviewed science, customer-visible commercial offerings, public price anchors, and public roadmaps. Third, it compares the major commercial platform families in terms that matter to a buyer, including access, operational burden, and upgrade risk. Fourth, it interprets public roadmaps as signals of refresh pressure rather than proof of present-day performance \citep{IBMRoadmap2026,QuantinuumRoadmap2024,PasqalRoadmap2025,QuandelaRoadmap2026}. The goal is not to rank vendors or predict a technological winner. It is to help institutions avoid the wrong first purchase. 

\begin{table}[t]
\centering
\small
\caption{Capability layers hidden inside the phrase ``buy a quantum computer.''}
\begin{tabularx}{\linewidth}{L{2.55cm}L{3.05cm}Y Y}
\toprule
Layer & What is acquired & When it fits & Representative public examples \\
\midrule
Preparation or deferral & Benchmark problems, staff time, classical baselines, and use-case validation instead of hardware & The workload is unclear, the user base is immature, or local ownership has no clear advantage & Internal benchmarking, cloud pilots, and classical--quantum workflow development \\
Multi-vendor cloud access & Execution time, simulators, notebooks, and basic support & Early exploration, first courses, small pilots, and platform discovery & IBM Open and pay-as-you-go plans; Amazon Braket cloud access \citep{IBMPlans2026,BraketPricing2026} \\
Reserved or premium access & Priority queueing, reserved capacity, and deeper technical support & Teams know what they want to run and need predictable throughput before ownership & QuEra Premium Access; D-Wave Leap and LaunchPad; Quantinuum cloud access \citep{QuEraPremium2026,DWaveLeap2026,DWaveLaunchPad2025,QuantinuumHelios2025} \\
Modest local instrument & A local system used often for teaching, controls, calibration, and methods development & The institution needs repeated local use and wants to build hardware practice without starting with a flagship machine & IQM Spark; Rigetti Novera; AQT rack-mounted trapped-ion systems \citep{IQMSpark2023,RigettiNovera2023,AQTSystems2025} \\
Strategic local installation & A large local platform with service, facilities, security, and upgrade planning & The mission requires sovereignty, tight hybrid integration, or sustained local throughput at organizational scale & Quantinuum Helios, QuEra on-premises systems, Atom AC1000, D-Wave Advantage2, OQC deployments, Quandela MosaiQ, ORCA PT-2 \citep{QuantinuumHelios2025,QuEraOnPrem2026,AtomAC1000_2026,DWaveSystems2026,OQCDeployments2026,QuandelaMosaiQ2026,ORCAPT2_2024} \\
\bottomrule
\end{tabularx}
\label{tab:layers}
\end{table}

\section{A procurement framework for quantum capability}

A disciplined procurement process begins by defining what the institution is actually trying to acquire. In quantum computing, the answer is rarely ``the most qubits.'' It is usually one of a smaller number of operational goals: broad access for students, repeated hands-on use for hardware research, secure and low-latency integration with classical infrastructure, or evidence that a target workload can benefit from quantum resources. Table~\ref{tab:layers} organizes these goals into five capability layers, each of which corresponds to a different operating model and a different institutional burden.

Role comes first because each layer solves a different problem. Preparation or deferral is a real procurement choice, not a failure to choose; it buys time to define benchmark tasks, train staff, and establish classical baselines. Multi-vendor cloud access is an exploration layer. It buys breadth, low initial cost, and exposure to more than one software stack. Reserved or premium access is a throughput layer. It buys predictable queue quality, user support, and a more stable path for courses, grant milestones, and early production-style workflows. A modest local instrument is a methods and culture layer. It buys repeated hands-on use, control experience, calibration practice, and local organizational learning. A strategic local installation is infrastructure. It buys sovereignty, security, and continuous availability, but it also commits the institution to facilities work, staffing, service contracts, and lifecycle planning \citep{IBMPlans2026,IBMFlexPlan2025,BraketPricing2026,QuEraPremium2026,DWaveLaunchPad2025,DWaveLeap2026,IQMSpark2023,RigettiNovera2023,AQTSystems2025,QuantinuumHelios2025,QuEraOnPrem2026,AtomAC1000_2026,DWaveSystems2026,OQCDeployments2026,QuandelaMosaiQ2026,ORCAPT2_2024}.

The mission usually follows institutional type. A teaching-focused campus cares about the number of users served per dollar, course reliability, and simple tooling. A research-intensive university may care about methods development, graduate training, and the ability to experiment repeatedly on local hardware. A national laboratory or sovereign HPC center may care most about data handling, co-scheduling with classical resources, and operational control. An industrial team may care almost exclusively about whether a benchmark workflow improves relative to the classical baseline. These missions overlap, but they should not be forced into the same purchase logic.

\begin{figure}[t]
\centering
\includegraphics[width=\linewidth]{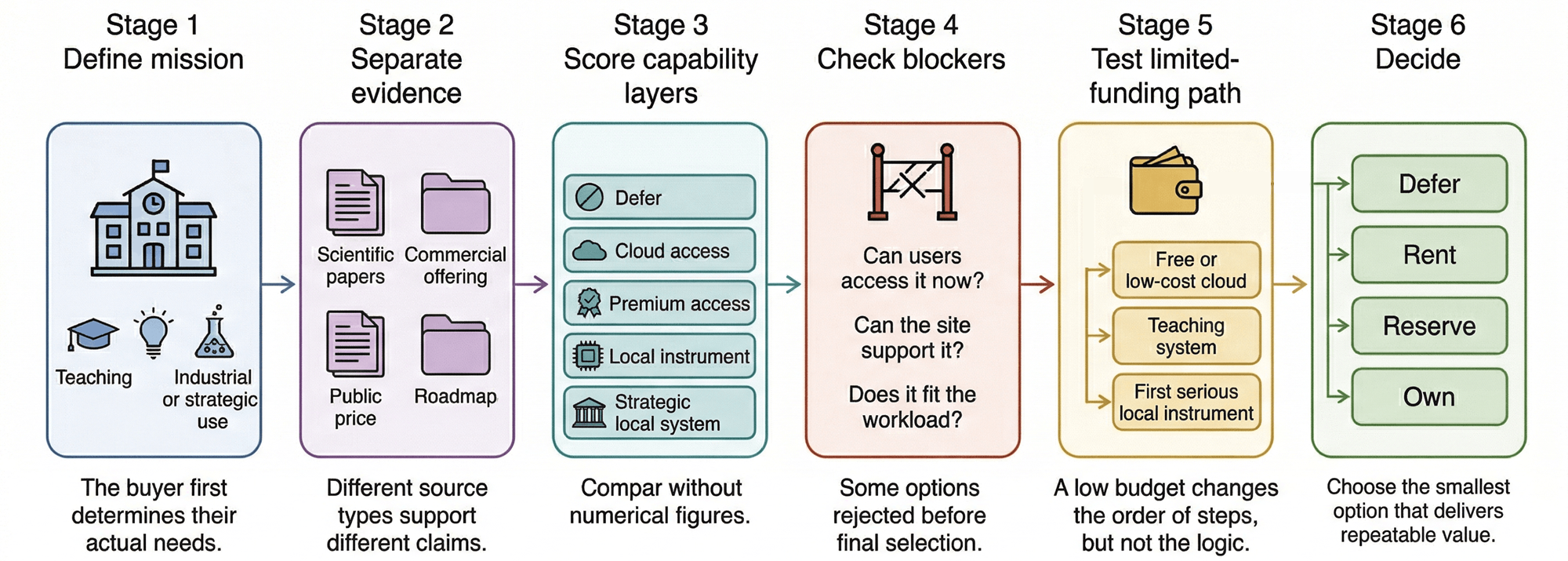}
\caption{\textbf{Procurement workflow.} A disciplined buying process starts with mission definition, separates evidence classes, scores capability layers, rejects blocked options, tests the low-budget path, and only then decides whether to defer, rent, reserve, or own.}
\label{fig:workflow}
\end{figure}

Figure~\ref{fig:workflow} turns the framework into a stage-gated process. Stage~1 defines the mission and the primary users. Stage~2 separates evidence types so that papers, product pages, prices, and roadmaps are not treated as interchangeable. Stage~3 scores capability layers without pretending to know more than the buyer actually knows. Stage~4 rejects blocked options, including systems the site cannot support or platforms that do not fit the workload. Stage~5 tests the low-budget path. Stage~6 makes the final decision only after the institution has made its operational assumptions explicit. This structure matters because many quantum procurements fail before the hardware arrives: the workload is vague, the user base is unprepared, or the support model is missing.

The evidence split is the most important control in the framework. Scientific papers show what a platform has demonstrated under research conditions. Commercial pages show what a customer can access now. Public price pages and procurement disclosures give rough spending scale. Roadmaps show where a company says it intends to go next. These are useful forms of evidence, but they support different claims. A paper showing progress in error correction is not evidence of a supported product. A product page promising access is not evidence of long-term technical leadership. A roadmap is not evidence of anything delivered today. Figure~\ref{fig:platforms} represents this distinction graphically by separating solid blocks, which indicate customer-visible offerings now, from dashed blocks, which indicate stated future targets only. Table~\ref{tab:roadmaps} applies the same discipline in words.

A simple weighted score is sufficient for first-pass comparison:
\begin{equation}
S_j = w_1R_{\mathrm{science},j} + w_2R_{\mathrm{operations},j} + w_3R_{\mathrm{access},j} + w_4R_{\mathrm{fit},j},
\end{equation}
where $R_{\mathrm{science}}$ measures technical fit to the workload, $R_{\mathrm{operations}}$ measures facilities and staffing burden, $R_{\mathrm{access}}$ measures how reliably users can run work now, and $R_{\mathrm{fit}}$ measures institutional fit across software, data handling, training, governance, and upgrade path. In practice, ordinal scores are usually enough. More numerical precision rarely improves the decision because much of the uncertainty is contractual and operational rather than mathematical. The buyer, not the vendor, chooses the weights. A teaching campus and a sovereign HPC center should not use the same weights because they are not buying the same thing.

Uncertainty should remain visible throughout the process. If pricing is quote-only, the model should widen the cost range instead of pretending that the price is known. If the service model is private, the framework should record operational uncertainty rather than assume support is weak. If a feature appears only on a roadmap, the buyer should treat it as unavailable unless it is written into the contract. Sensitivity analysis is also useful. If a recommendation changes whenever one weight is nudged slightly, the institution does not yet have a robust decision. Table~\ref{tab:questions} later translates these uncertainties into due-diligence questions that belong in the procurement package, not in footnotes.

A sound process can also conclude that the right purchase today is no purchase at all. If internal users cannot identify benchmark problems, success metrics, ownership responsibilities, and the reason local ownership would be superior to access, then a cloud pilot or funded preparation period is often the highest-value outcome. In a fast-moving market, disciplined deferral can be a strong decision.

\section{Platforms, companies, and the road ahead}

Platform choice affects more than raw performance. It also shapes the control stack, software environment, staffing burden, site requirements, and likely refresh cycle. From a buyer's perspective, platform choice is therefore a decision about operations as much as about qubit physics. Figure~\ref{fig:platforms} summarizes the field by platform family and separates systems sold now from roadmap-only claims. Table~\ref{tab:roadmaps} expands that picture into a vendor-by-vendor procurement reading. Together they show why a platform label is not enough: the buying experience inside one platform family can differ sharply from company to company.  {For procurement, the most useful scientific question is not only what vendors promise, but what the platform has already achieved in peer-reviewed work. That literature is the clearest public evidence of what a well-prepared institution may realistically be able to do with the hardware family during the lifetime of a purchase.}

\subsection{How to read public roadmaps}

Public roadmaps are useful, but they answer a narrow question. They indicate how quickly the commercial frontier may move after a contract is signed. They do not tell the buyer what is delivered now, what level of support is included, or how the system performs on the institution's target workloads. This is why the separation between solid and dashed blocks in Figure~\ref{fig:platforms} matters. Solid blocks mark present customer-visible offerings. Dashed blocks mark future targets only. Table~\ref{tab:roadmaps} follows the same rule in text and converts public plans into a practical procurement reading \citep{IBMRoadmap2026,QuantinuumRoadmap2024,PasqalRoadmap2025,QuandelaRoadmap2026,DWaveGateRoadmap2026}.

Roadmaps also need careful vocabulary. A \textit{physical qubit} is the hardware element the device directly controls. A \textit{logical qubit} is an error-protected qubit encoded across many physical qubits. \textit{Fault tolerance} means the machine can keep logical error rates under control as computations become deeper and longer. These definitions matter because vendors report progress using different quantities. Physical qubit counts, logical qubit counts, application benchmarks, and company-specific metrics are not interchangeable. IonQ, for example, prominently uses algorithmic qubits, a company-defined measure of usable system performance, alongside physical and logical qubit milestones \citep{IonQRoadmap2025}. The buyer should therefore read every roadmap against the delivery model, not in isolation from it.

 {A second procurement check is to compare every roadmap with a platform's best peer-reviewed demonstrations. Roadmaps show where a vendor hopes to go. Peer-reviewed studies show what scientists have already made the hardware family do under demanding conditions. The two evidence classes answer different questions, and a serious buyer should read both.}

The practical rule is simple: roadmaps are procurement signals, not procurement evidence. They matter because they compress the likely useful lifetime of a local purchase and affect how aggressively a buyer should negotiate upgrade rights, refresh options, acceptance tests, and service commitments. They should not be treated as proof that the next milestone will arrive on schedule.

\subsection{Superconducting gate-model systems}

Superconducting systems remain the broadest commercial ecosystem. These machines use superconducting electrical circuits as qubits and typically operate at very low temperatures inside dilution refrigerators. Their advantages include fast gates, strong fabrication ecosystems, and mature software toolchains. Their trade-offs include cryogenic infrastructure, calibration burden, and substantial control complexity \citep{Siddiqi2021}. For procurement, superconducting systems are often the easiest platform family to access broadly and among the most demanding to own locally at scale.

 {Peer-reviewed milestones already show what this platform can do at the frontier. On IBM hardware, error-mitigated experiments on a 127-qubit superconducting processor measured accurate expectation values for circuits with up to 60 layers of two-qubit gates and 2,880 CNOT gates, providing evidence for pre-fault-tolerant utility in many-body dynamics \citep{Kim2023Utility}. Google's superconducting processors first showed that logical performance can improve as surface-code distance increases from $d=3$ to $d=5$ \citep{GoogleSurfaceCode2023}, and then reached below-threshold operation with a 101-qubit $d=7$ surface-code memory with $0.143\%\pm0.003\%$ logical error per cycle and real-time decoding \citep{GoogleBelowThreshold2025}. For a buyer, these demonstrations show that superconducting systems already support both ambitious NISQ-scale quantum simulation and early logical-qubit experiments.}

IBM anchors this platform family because it offers the clearest access ladder. Buyers can start with a free Open Plan, move to paid access, and then negotiate deeper enterprise arrangements. IBM also publishes one of the clearest public roadmaps in the field. Its current public hardware roadmap targets near-term quantum advantage by 2026 and a first large-scale fault-tolerant quantum computer by 2029 \citep{IBMPlans2026,IBMRoadmap2026,IBMHardware2026}. For procurement, that roadmap is best read as a refresh-pressure signal. It suggests that the superconducting frontier may move on a timescale shorter than many capital-equipment cycles, which strengthens the case for caution when considering first-generation local ownership.

Rigetti, IQM, and OQC show that ``superconducting'' is not one buying style. Rigetti offers the Novera quantum processing unit as a small on-premises research entry point, and it has advanced a modular chiplet path from the 36-qubit Cepheus-1-36Q system toward a 108-qubit Cepheus system \citep{RigettiNovera2023,RigettiCepheus36Q2025,RigettiCepheus108Q2026}. IQM has built a staged on-premises ladder that is unusually relevant for universities and national labs: Spark for education and first local use, Radiance for larger deployments, and Halocene for quantum error-correction development. IQM's public roadmap targets fault-tolerant computing by 2030 and scale toward one million qubits \citep{IQMSpark2023,IQMRadiance2023,IQMRoadmap2030,IQMHalocene2025}. OQC tells a different story again. Its public material emphasizes direct access, colocation, and deployment in commercial and sovereign data centers, alongside a roadmap that moves from GENESIS to TITAN, ATHENA, and ATLAS \citep{OQCToshiko2023,OQCRoadmap2026,OQCDirect2026,OQCDeployments2026}. The lesson is that the platform label alone does not determine the buyer experience; access and deployment model matter just as much.

Google matters here as a technical reference point even though it is not a routine procurement vendor. Its public roadmap still centers on a staged path to a useful, error-corrected quantum computer, and its recent superconducting error-correction results continue to shape expectations across the field \citep{GoogleRoadmap2026,GoogleSurfaceCode2023,GoogleBelowThreshold2025}. That influence matters because many buyers measure commercial announcements against Google-style milestones even when Google is not offering them a product.

\subsection{Trapped-ion systems}

Trapped-ion systems use individual atomic ions as qubits and control them with lasers or microwave fields. Their strongest commercial case usually appears when buyers value long coherent operations, high-fidelity control, flexible connectivity, and logical-qubit development more than raw user scale \citep{Bruzewicz2019}. Their trade-offs include slower gate speeds than superconducting circuits and substantial optics or laser-system complexity. From a procurement perspective, trapped ions often appeal less as a mass-access teaching platform and more as a premium research resource.

 {The trapped-ion literature already shows a wide span of demonstrated capability. On Quantinuum's H2 processor, researchers created the ground-state wavefunction of non-Abelian $D_{4}$ topological order on a 27-qubit kagome lattice with fidelity per site above $98.4\%$, a demanding test of coherent control and many-body state preparation \citep{Iqbal2024Anyons}. The same hardware family also enabled high-fidelity teleportation of a logical qubit using transversal gates and lattice surgery, demonstrating that trapped-ion systems are relevant not only for small algorithms, but also for logical operations and modular fault-tolerant workflows \citep{RyanAnderson2024Teleportation}. These demonstrations matter in procurement terms because they show what institutions can realistically aim to do on premium trapped-ion hardware beyond routine benchmark circuits.}

IonQ publishes one of the boldest numerical public roadmaps among commercial vendors. The company states a path to 2 million physical qubits and 80{,}000 logical qubits by 2030, and it has positioned Forte Enterprise and Tempo as enterprise-facing systems for production readiness and larger commercial workloads \citep{IonQForteEnterprise2026,IonQTempo2025,IonQRoadmap2025}. This clarity helps procurement because it exposes the vendor's timing assumptions. It does not remove the need for diligence on present access, service commitments, or delivered performance.

Quantinuum offers a more premium trapped-ion path. Helios is available through cloud and on-premises access, and the company's public roadmap places Apollo as a universal, fully fault-tolerant system at the end of the decade. Quantinuum has also extended that roadmap into the 2030s with Lumos and linked the longer-term story to the Defense Advanced Research Projects Agency's utility-scale benchmarking agenda \citep{QuantinuumHelios2025,QuantinuumRoadmap2024,QuantinuumDARPA2025}. This positioning makes Quantinuum especially relevant for buyers who care about logical depth, chemistry, and long-horizon fault-tolerant research.

AQT occupies a different trapped-ion niche. Its public offer emphasizes room-temperature operation, a compact rack-mounted footprint, cloud access, and integration with existing computing environments. That profile makes AQT attractive when facilities simplicity and local operability matter as much as frontier branding \citep{AQTSystems2025,AQTBraket2025,AQTPoland2024}. In other words, trapped ions span at least three buyer personas: premium logical-computing access, enterprise system positioning, and compact local deployment.

\subsection{Neutral-atom systems}

Neutral-atom systems trap and move neutral atoms with laser light and use reconfigurable atom arrays to define qubits and interactions. This platform has gained momentum because it combines large arrays, flexible connectivity, analog and early digital operating modes, and rapid progress in error-correction research \citep{Wintersperger2023,Bluvstein2025FT}. It is also a platform family in which the distinction between what can be run today and what is promised for logical-computing roadmaps matters especially strongly.

 {Neutral atoms now offer some of the clearest peer-reviewed evidence of rapid movement toward logical computing. A programmable logical processor based on reconfigurable atom arrays operated with up to 280 physical qubits, improved two-qubit logical gates by scaling surface-code distance from $d=3$ to $d=7$, created fault-tolerant logical GHZ states, performed feedforward entanglement teleportation, and executed 48-logical-qubit sampling circuits with 228 logical two-qubit gates and 48 logical CCZ gates \citep{Bluvstein2024LogicalProcessor}. More recently, neutral-atom machines demonstrated logical magic-state distillation \citep{SalesRodriguez2025MagicStateDistillation} and the key elements of a universal fault-tolerant architecture using up to 448 atoms, including $2.14(13)\times$ below-threshold performance in a four-round characterization circuit \citep{Bluvstein2026FTArchitecture}. For procurement, these results are unusually important because they show that neutral atoms are no longer only large-array simulators; they are becoming serious platforms for logical-qubit research.}

QuEra's public story has shifted quickly. The company's 2024 roadmap laid out a path to 30 logical qubits in 2025 and 100 logical qubits in 2026. Its current roadmap language says the program has progressed faster than expected and now frames full fault tolerance as a 2029 goal. At the same time, QuEra sells several access modes: public cloud access, Premium Access with direct scientist support, and on-premises deployment for secure, continuous use \citep{QuEraRoadmap2024,QuEraRoadmap2026,QuEraPremium2026,QuEraOnPrem2026}. Buyers should read this combination carefully. QuEra is both a practical learning platform today and a roadmap-heavy logical-computing bet.

Pasqal offers the clearest upgradable neutral-atom product ladder now in the public domain. Its 2025 roadmap states a path from 2 logical qubits in 2025 to 20 in 2027, 100 high-fidelity logical qubits in 2029, and 200 in 2030. On the hardware side, the company links Orion, Vela, Centaurus, and Lyra to that progression and pairs the hardware with a software stack aimed at hybrid industrial workflows \citep{PasqalRoadmap2025,PasqalHardware2026,PasqalSoftware2026}. Pasqal's deployment at CINECA also shows how the company is positioning itself for hybrid HPC integration \citep{PasqalCINECA2026}. For buyers who want an explicit path from early access to larger integrated deployment, that coherence matters.

Atom Computing occupies the large-infrastructure end of the neutral-atom market. Its AC1000 system is marketed with more than 1{,}200 physical qubits and a feature set intended to support quantum error correction, including mid-circuit measurement, reset, reuse, and all-to-all connectivity. Atom's technology pages also stress an architectural path to thousands and then millions of qubits \citep{AtomAC1000_2026,AtomTechnology2026,AtomHome2026}. For procurement, the message is clear: Atom is selling scale, logical-qubit development, and on-premises ambition rather than a small first-step instrument.

\subsection{Quantum annealing systems}

Quantum annealing deserves its own procurement logic. It maps optimization problems onto an energy landscape and searches for low-energy solutions. It is not a general gate-model approach, and buyers should not treat it as a substitute for one. At the same time, it is not a lesser version of universal quantum computing. It is a specialized computing model whose value depends on whether the target workload already looks like a good optimization or sampling problem \citep{DWaveSystems2026}.

 {Peer-reviewed annealing results also clarify what institutions can realistically expect. A 5,000-qubit programmable spin glass on D-Wave hardware reproduced quantum-critical dynamics on thousands of qubits and extracted critical exponents that distinguished quantum annealing from slower stochastic classical dynamics \citep{King2023QuantumCritical}. A later \emph{Science} study showed that superconducting quantum annealers can generate samples in close agreement with Schr\"{o}dinger-equation solutions on a quantum simulation task, establishing beyond-classical performance for that problem class \citep{King2025BeyondClassical}. These results do not make annealers universal machines, but they do show that annealing hardware can already be scientifically serious when the workload is naturally phrased as optimization, sampling, or many-body simulation.}

D-Wave remains the clearest commercial case for mission fit over prestige. The company sells the Advantage2 annealing system through cloud access and on-premises deployment, and it supports enterprise entry through Leap and LaunchPad \citep{DWaveSystems2026,DWaveLeap2026,DWaveLaunchPad2025}. For institutions with concrete optimization pipelines, this matters more than abstract long-term comparisons with universal machines. In many settings, the right procurement question is not ``Will annealing win the platform race?'' but ``Can this workflow produce measurable value now?''

D-Wave has also widened its public story. After acquiring Quantum Circuits, the company published a three-year gate-model roadmap that targets general availability of a 17-qubit dual-rail system in 2026, a 49-qubit system in 2027, and a 181-qubit system in 2028, together with multiple-logical-qubit error-correction demonstrations and a scalable 1{,}000-qubit processor design. It has also reported scalable on-chip cryogenic control of gate-model qubits \citep{DWaveAcquisition2026,DWaveGateRoadmap2026,DWaveCryoControl2026}. This does not erase D-Wave's identity as the leading annealing vendor. It does mean that the company should now be read as a dual-platform supplier with a strong application-driven installed base and an emerging gate-model narrative.

\vspace{10pt}

\begingroup
\footnotesize
\begin{longtable}{L{1.85cm}L{2.45cm}L{4.65cm}L{3.85cm}L{1.55cm}}
\caption{Major commercial platforms and public roadmaps. The table summarizes buyer-visible offerings and public disclosures. It is not a universal technology ranking.}\\
\toprule
Vendor & Buyer-facing position now & Public roadmap or disclosed path & Practical reading for a buyer & Key sources \\
\midrule
\endfirsthead
\toprule
Vendor & Buyer-facing position now & Public roadmap or disclosed path & Practical reading for a buyer & Key sources \\
\midrule
\endhead
IBM & Broadest public access ladder in superconducting systems & Public hardware roadmap targets near-term quantum advantage by 2026 and a first large-scale fault-tolerant machine by 2029 & The default broad-access benchmark and one of the clearest signals of refresh pressure & \citep{IBMPlans2026,IBMRoadmap2026,IBMHardware2026} \\
Rigetti & Small on-premises research hardware and modular multi-chip superconducting systems & Public path runs from Novera and the 36-qubit Cepheus-1-36Q to the 108-qubit Cepheus system & Best read as a local superconducting research path rather than a campus-wide utility layer & \citep{RigettiNovera2023,RigettiCepheus36Q2025,RigettiCepheus108Q2026} \\
IQM & Staged on-premises ladder from Spark to Radiance to Halocene & Public roadmap targets fault-tolerant computing by 2030; Halocene targets error-correction users while Radiance serves larger HPC deployments & One of the clearest paths from a first local instrument to an error-correction program & \citep{IQMSpark2023,IQMRadiance2023,IQMRoadmap2030,IQMHalocene2025} \\
OQC & Secure direct access and deployment in commercial and sovereign data centers & Public roadmap moves from GENESIS to TITAN, ATHENA, and ATLAS & Strong fit where deployment model and data-center presence matter as much as headline qubit count & \citep{OQCToshiko2023,OQCRoadmap2026,OQCDirect2026,OQCDeployments2026} \\
IonQ & Enterprise trapped-ion systems and cloud access & Public roadmap states milestones through 2030 and uses physical, logical, and algorithmic qubit metrics & Clear long-range planning signal, but current buyer diligence still matters more than the curve on the slide & \citep{IonQForteEnterprise2026,IonQTempo2025,IonQRoadmap2025} \\
Quantinuum & Premium trapped-ion access with cloud and on-premises Helios & Public roadmap places Apollo at the end of the decade and extends into the 2030s with Lumos & Strongest premium logical-qubit story in the current market & \citep{QuantinuumHelios2025,QuantinuumRoadmap2024,QuantinuumDARPA2025} \\
AQT & Compact trapped-ion systems with cloud and local deployment options & Public product story emphasizes room-temperature, rack-mounted operation more than a long dated roadmap & Attractive where facilities simplicity and local operability dominate the buying decision & \citep{AQTSystems2025,AQTBraket2025} \\
QuEra & Neutral-atom cloud, premium, and on-premises systems & 2024 roadmap stated 30 logical qubits in 2025 and 100 in 2026; current roadmap frames full fault tolerance as a 2029 target & Fast-moving roadmap with real access options now; buyers should separate present systems from future logical targets & \citep{QuEraRoadmap2024,QuEraRoadmap2026,QuEraPremium2026,QuEraOnPrem2026} \\
Pasqal & Neutral-atom systems for cloud and on-premises HPC-linked deployment & 2025 roadmap states a path from 2 logical qubits in 2025 to 200 in 2030 and links Orion, Vela, Centaurus, and Lyra to that path & The clearest upgradable neutral-atom commercial roadmap now public & \citep{PasqalRoadmap2025,PasqalHardware2026,PasqalCINECA2026} \\
Atom Computing & Large neutral-atom infrastructure focused on scale and logical qubits & Public material stresses 1{,}200+ qubits now and a path to thousands and millions of qubits & Strong scale story for buyers already thinking about local strategic infrastructure & \citep{AtomAC1000_2026,AtomTechnology2026,AtomHome2026} \\
D-Wave & Commercial annealing systems plus a new gate-model program & Public material combines Advantage2 with Leap and LaunchPad access and a three-year dual-rail gate-model roadmap & Best current case where optimization value exists now and platform diversification is public & \citep{DWaveSystems2026,DWaveLeap2026,DWaveLaunchPad2025,DWaveGateRoadmap2026} \\
Quandela & Modular photonic systems with cloud and on-premises deployment & Public 2026--2033 roadmap links near-term processors to a logical-era ladder and utility-scale targets & The clearest dated commercial roadmap in photonics & \citep{QuandelaMosaiQ2026,QuandelaBelenos2025,QuandelaRoadmap2026} \\
ORCA & Rack-mounted photonic systems for quantum-enhanced machine learning and hybrid integration & Public deployment story is clear; long-range public roadmap is less explicit than Quandela's & Evaluate by application fit and infrastructure model, not by absent headline milestones & \citep{ORCAPT2_2024,ORCATechnology2026,ORCANQCC2025} \\
\bottomrule
\label{tab:roadmaps}
\end{longtable}
\endgroup

\subsection{Photonic systems}

Photonic systems use photons as quantum carriers. Their strongest commercial claims usually center on room-temperature operation for many components, modular networking, and compatibility with fiber and data-center infrastructure. For buyers, the key question is whether a vendor is offering a path toward general-purpose fault-tolerant computing, a special-purpose accelerator, or some mixture of both \citep{AghaeeRad2025}. In procurement terms, photonics is not one market story.

 {Photonics already spans three distinct peer-reviewed capability levels. First, programmable photonic processors have demonstrated quantum computational advantage in Gaussian boson sampling on 216 squeezed modes \citep{Madsen2022QuantumAdvantage}. Second, Quandela's cloud-accessible single-photon platform benchmarked one-, two-, and three-qubit gates at 99.6\%, 93.8\%, and 86\%, respectively, and ran a variational quantum eigensolver for the hydrogen molecule with chemical accuracy \citep{Maring2024VersatilePhotonicPlatform}. Third, the platform now has explicit fault-tolerant building blocks: GKP logical states have been generated in propagating light \citep{Konno2024LogicalStates}, modular fibre-networked photonic architectures built from 35 chips have demonstrated the primitive functions of a universal fault-tolerant photonic computer \citep{AghaeeRad2025ModularPhotonic}, and manufacturable silicon-photonics modules have reported 99.22\% two-qubit fusion fidelity and 99.72\% chip-to-chip qubit interconnect fidelity \citep{PsiQuantum2025Manufacturable}. Together, these results show that photonics already covers both near-term algorithm demonstrations and long-horizon fault-tolerant building blocks.}

Quandela currently offers the clearest public photonic commercial roadmap. Its MosaiQ platform is sold as a modular, datacenter-ready photonic computer, and its 2026--2033 roadmap lays out a progression from near-term acceleration pilots to the fault-tolerant regime and then large-scale quantum utility. The same public materials connect that roadmap to a product ladder that includes Belenos, Canopus, Andromeda, Sirius, and Ursa Major \citep{QuandelaMosaiQ2026,QuandelaBelenos2025,QuandelaRoadmap2026}. For procurement, Quandela is the strongest current example of a photonic vendor that links a present commercial product to a dated logical-era story.

ORCA represents a different photonic buying model. Its PT-2 is a rack-mounted, room-temperature photonic system aimed at quantum-enhanced machine learning and HPC integration. ORCA also emphasizes telecom-grade components and deployment into existing infrastructure, and it has delivered a system to the United Kingdom's National Quantum Computing Centre \citep{ORCAPT2_2024,ORCATechnology2026,ORCAHome2026,ORCANQCC2025}. That profile makes ORCA more natural for buyers who want an operational accelerator story now than for buyers who want a detailed dated roadmap to thousands of logical qubits. More broadly, photonics already spans both long-horizon general-purpose ambition and near-term application-shaped deployment.

\subsection{What the roadmap means for buyers}

Table~\ref{tab:roadmaps} should not be read as a leaderboard. It is a map of buyer choices and refresh pressure. IBM and Amazon Braket define broad-access baselines. Quantinuum and IonQ define premium trapped-ion trajectories. QuEra, Pasqal, and Atom define very different neutral-atom buying styles. D-Wave remains the clearest fit-for-optimization option while also entering the gate-model discussion. Quandela and ORCA show that photonics already spans both general-purpose ambition and application-driven deployment.

 {The practical lesson is that a roadmap should always be filtered through demonstrated science. A platform with peer-reviewed evidence of logical teleportation, below-threshold memories, magic-state distillation, or large-scale many-body simulation belongs in a different procurement class from one that offers only a persuasive future slide. Buyers should therefore treat the literature as a capability floor and the roadmap as a capability hypothesis.}

The practical consequence is direct. A public roadmap can shorten the sensible life of a local purchase even when the current system is good. Buyers should therefore negotiate upgrade rights, acceptance tests, service commitments, and training deliverables with Table~\ref{tab:questions} in hand. Roadmaps do not settle who will win. They do tell you how expensive it may be to buy a snapshot instead of a platform.

\begin{figure}[t]
\centering
\includegraphics[width=\linewidth]{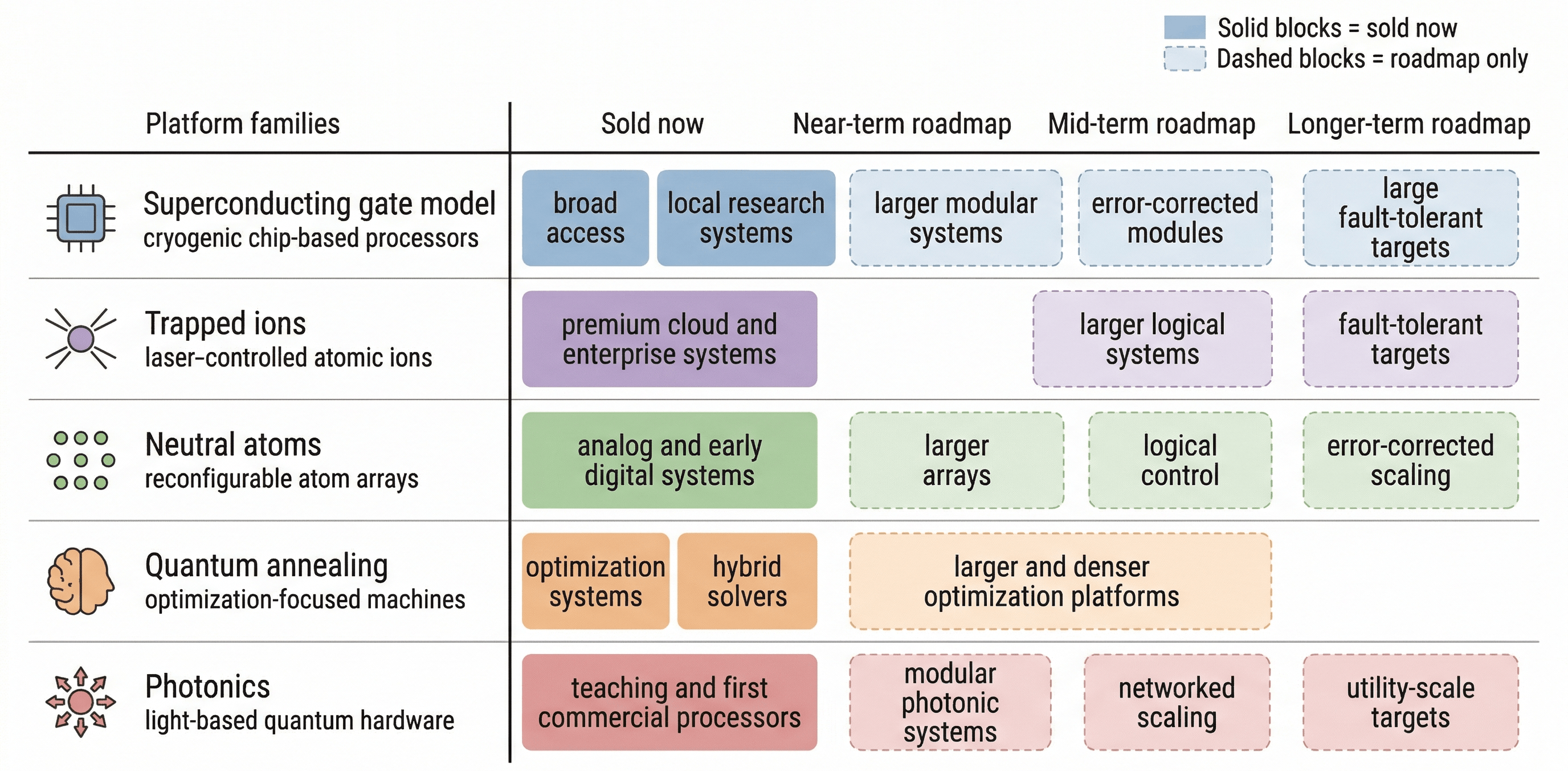}
\caption{\textbf{Platforms and public roadmaps.} Solid blocks indicate customer-visible offerings now. Dashed blocks indicate public roadmap targets only. The figure should be read as a market map and a refresh-pressure signal, not as a cross-platform performance ranking.}
\label{fig:platforms}
\end{figure}

\section{Where to start when funding is limited}

\begin{table}[t]
\centering
\footnotesize
\caption{Where to start when funding is limited. The table separates access tools, instructional systems, and the first serious local instruments.}
\begin{tabularx}{\linewidth}{L{2.65cm}L{2.35cm}L{2.45cm}Y}
\toprule
Option & Type & Public cost posture & Best first use \\
\midrule
IBM Quantum Open and paid plans & Cloud access & Free entry plus paid usage tiers & First courses, small research pilots, and early gate-model workflows \\
Amazon Braket & Multi-vendor cloud access & No local capital spend for first exploration & Comparing modalities and building internal benchmark suites \\
D-Wave Leap and LaunchPad & Annealing and hybrid cloud access & Trial and guided entry path & Optimization pilots before any local commitment \\
Thorlabs and qutools teaching platforms & Instructional photonics & Laboratory equipment rather than compute infrastructure & Quantum optics, entanglement labs, and communication concepts \\
SpinQ Gemini Mini and Gemini Lab & Small real-hardware teaching systems & Educational hardware with limited scale & Hands-on instruction and laboratory culture building \\
IQM Spark & First serious local instrument & Affordable on-premises entry point for universities and labs & Repeated local use, methods research, and hybrid workflow training \\
Rigetti Novera and AQT systems & First serious local research instruments & Quote-based local systems & Calibration practice, controls work, and focused quantum methods research \\
\bottomrule
\end{tabularx}
\label{tab:limited}
\end{table}

Limited funding should change the sequence of purchases, not stop the program. The right early deliverable is rarely a machine. It is a portfolio of evidence: benchmark problems, trained users, reproducible workflows, a clearer view of software portability, and a better understanding of what local ownership would actually add. Table~\ref{tab:limited} lays out the sensible order. The pattern is simple: start with access, then add teaching hardware if education is a core mission, then buy a first serious local instrument only when repeated local use is already visible. Figure~\ref{fig:workflow} places this logic in Stage~5, where the buyer tests the low-budget path before committing to ownership.

Cloud access remains the strongest starting point for most institutions. IBM and Amazon Braket lower the barrier to entry because they let users run real hardware and simulators without building local infrastructure. D-Wave's Leap and LaunchPad do the same for annealing and hybrid optimization workflows \citep{IBMPlans2026,BraketPricing2026,DWaveLeap2026,DWaveLaunchPad2025}. For most early programs, this cloud-first stage buys the most valuable asset: evidence. It also lets institutions compare interfaces, queue behavior, and workflow portability before they become dependent on a single stack.

A multi-vendor approach is especially useful at the beginning. Using more than one cloud provider or hardware modality forces the organization to define what it actually needs instead of inheriting one vendor's assumptions. It also clarifies whether the real bottleneck is hardware access, algorithm design, hybrid workflow engineering, or internal training. That insight is often more valuable than early ownership.

Teaching hardware is the next step, but it should not be confused with computing hardware. Thorlabs and qutools sell photonics education systems that are excellent for demonstrating single-photon effects, entanglement, Bell tests, and quantum communication concepts. These systems create substantial educational value, but they do not substitute for a commercial gate-model, neutral-atom, trapped-ion, annealing, or photonic computing platform \citep{ThorlabsQOP2026,QutoolsQuED2026}. Institutions often make poor first purchases because they try to solve teaching, exploration, and infrastructure with one device.

Small real-hardware systems fill yet another role. SpinQ's Gemini Mini and Gemini Lab are designed for classroom and laboratory use, not for frontier-scale research computing. That limit is a virtue when the real goal is to build laboratory intuition, give students hands-on experience, and develop a local culture around quantum instrumentation \citep{SpinQMini2026,SpinQLab2026}. A modest educational system can be a good purchase even when it is a poor research computer.

The first serious local instrument comes later. IQM Spark is the clearest example because it is marketed as an affordable on-premises system for universities and labs. Rigetti Novera makes sense when cryogenic integration and superconducting controls are part of the learning goal. AQT makes sense when a buyer wants a trapped-ion platform with lower facilities friction. Table~\ref{tab:limited} keeps these roles separate because institutions often make poor decisions when they mix teaching, exploration, and infrastructure into one purchase \citep{IQMSpark2023,RigettiNovera2023,AQTSystems2025}.

There is also a middle path between cloud-only use and full ownership: shared regional infrastructure, consortium access, or co-funded deployment. Not every institution needs its own local system to gain repeated access. In many cases, the best first local presence is organizational rather than hardware-based: a shared benchmark suite, trained users, and a clear governance model.

\section{Cost, lock-in, and timing}

\begin{figure}[t]
\centering
\includegraphics[width=0.92\linewidth]{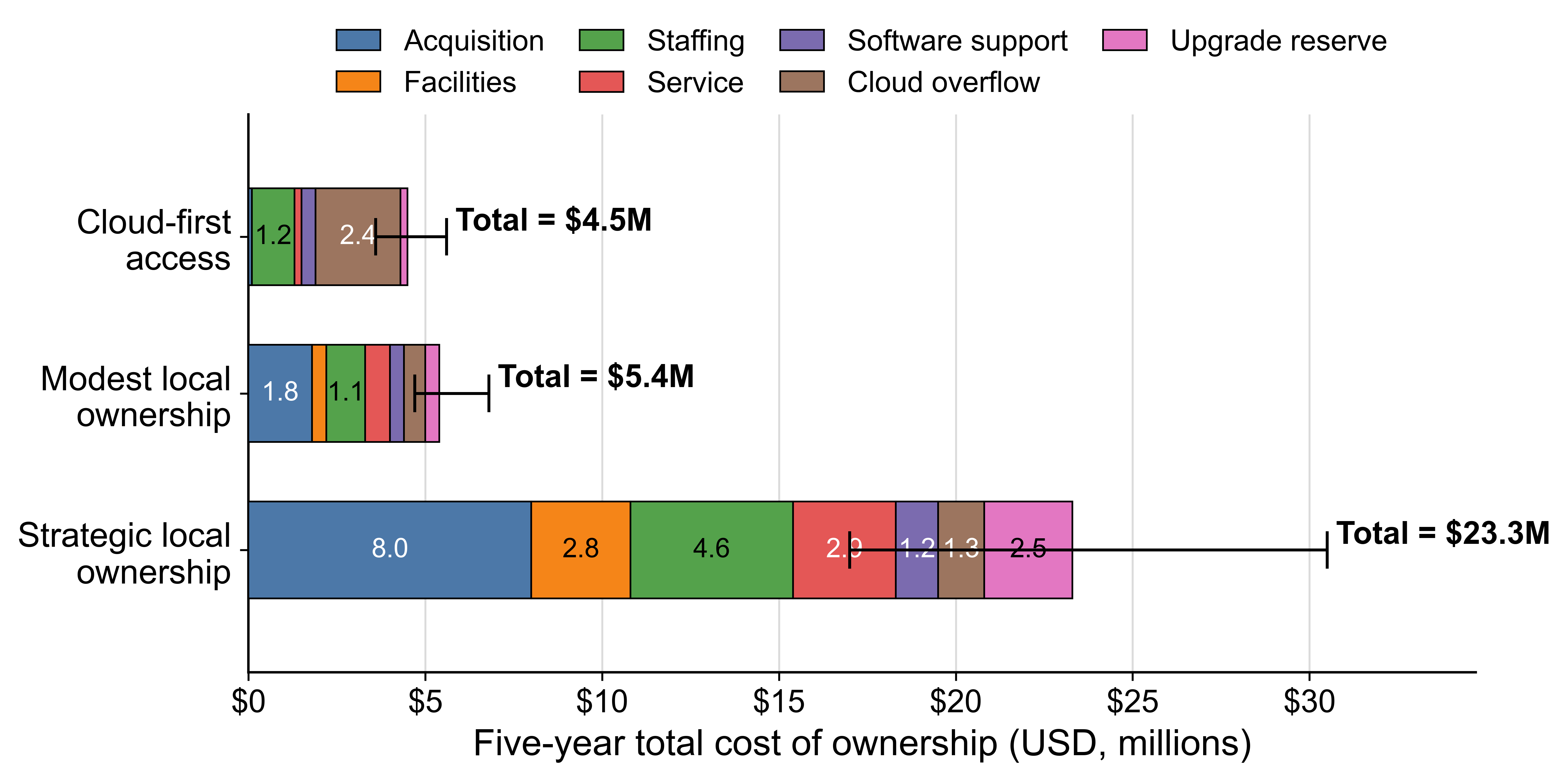}
\caption{\textbf{Illustrative five-year total cost of ownership.} The figure compares three operating models---cloud-first access, modest local ownership, and strategic local ownership---using explicit cost blocks for acquisition, facilities, staffing, service, software support, cloud overflow, and upgrade reserve. These values are scenario estimates for comparison, not vendor quotes.}
\label{fig:cost}
\end{figure}

Total cost of ownership matters more than sticker price. Figure~\ref{fig:cost} makes the point directly. In the illustrative five-year model shown there, cloud-first access stays below the cost of even modest ownership because it avoids acquisition, facilities work, and some recurring support. Modest local ownership remains manageable only when the buyer already has a use pattern that justifies repeated local use. Strategic local ownership becomes expensive because acquisition, staffing, service, and upgrade reserve rise together. The figure is illustrative, not predictive, but the ordering is the point: the operating model often dominates the hardware line item.

The budget is often misread because acquisition is the most visible cost and not necessarily the largest one. As soon as a system must operate as an institutional resource rather than as a single-lab instrument, recurring people costs grow quickly. Someone has to manage users, coordinate with the vendor, maintain software, oversee acceptance testing, and keep workflows running. Service contracts and software support then become structural costs rather than optional add-ons. Figure~\ref{fig:cost} should therefore be read together with Table~\ref{tab:questions}: the contract terms for maintenance, upgrades, training, and support are part of the cost model, not details to negotiate at the end.

\begin{table}[t]
\centering
\footnotesize
\caption{Questions that belong in the due-diligence package before any institutional quantum purchase is signed.}
\begin{tabularx}{\linewidth}{L{4.05cm}Y L{2.25cm}}
\toprule
Question & Why it matters & Representative sources \\
\midrule
What is delivered now, and what remains roadmap only? & It prevents future targets from entering the contract as if they were shipped features & \citep{IBMRoadmap2026,QuantinuumRoadmap2024,PasqalRoadmap2025,QuandelaRoadmap2026,DWaveGateRoadmap2026} \\
What are the real site requirements? & Floor space, power, cooling, integration, and security rules can block a project before qubit metrics matter & \citep{IQMSpark2023,QuEraOnPrem2026,DWaveSystems2026,AQTSystems2025,AtomTechnology2026,QuandelaMosaiQ2026,ORCATechnology2026} \\
Who owns calibration, maintenance, and acceptance testing after delivery? & The answer distinguishes access, appliance-like deployment, and research instrument ownership & \citep{RigettiNovera2023,IQMSpark2023,QuEraOnPrem2026,QuantinuumHelios2025} \\
What is upgradeable in place, and what requires replacement? & Upgrade rights determine whether the first purchase can evolve or whether it will age as a closed box & \citep{IQMRoadmap2030,IQMHalocene2025,PasqalRoadmap2025,OQCRoadmap2026,QuandelaRoadmap2026} \\
What software, training, and hybrid workflow support are included? & Time to first useful result depends as much on tooling and training as on hardware & \citep{IBMPlans2026,BraketPricing2026,PasqalSoftware2026,ThorlabsQOP2026,QutoolsQuED2026} \\
How are data handling, direct access, and HPC integration managed? & Security and hybrid integration are often the real reason to buy locally at all & \citep{OQCDirect2026,OQCDeployments2026,QuEraOnPrem2026,QuantinuumHelios2025,PasqalCINECA2026} \\
What metric defines successful delivery? & A clear acceptance test keeps the purchase tied to measurable value instead of vague future potential & \citep{IBMPlans2026,IQMSpark2023,QuEraOnPrem2026,DWaveSystems2026} \\
\bottomrule
\end{tabularx}
\label{tab:questions}
\end{table}

Queue quality is part of cost because users experience access, not qubit count. A strong machine with poor access wastes student time, delays deliverables, and can make a funded program look technically weak even when the hardware itself is not the problem. Premium access tiers exist because predictable throughput has real value for courses, proposal deadlines, and hybrid production workflows \citep{QuEraPremium2026,DWaveLeap2026,QuantinuumHelios2025}. Measured this way, reserved access is often a cheaper solution than ownership.

Lock-in is also part of cost. It appears at three levels. Workflow lock-in occurs when job submission, calibration interfaces, and hybrid solvers are tightly coupled to one proprietary environment. Data lock-in occurs when results, metadata, or benchmark traces are difficult to export or compare. Workforce lock-in occurs when training leaves the institution fluent in one stack but not in transferable methods. A team that builds all of its habits inside one closed environment may discover later that it bought switching costs instead of strategic leverage \citep{OQCDirect2026,PasqalSoftware2026,ORCATechnology2026}. This is why portability and API stability belong in the procurement discussion from the beginning.

Facilities are part of cost because buildings move slowly. Superconducting systems demand cryogenic infrastructure. Neutral-atom, trapped-ion, and photonic systems often have lighter utility demands, but they still need floor space, integration work, networking, cybersecurity review, environmental stability, and local support. On-premises pages from QuEra, Atom, OQC, Quandela, and others make clear that the system always lives somewhere physical, even when the sales pitch starts in the cloud \citep{QuEraOnPrem2026,AtomTechnology2026,OQCDeployments2026,QuandelaMosaiQ2026,ORCATechnology2026}. Site readiness can therefore be a first-order constraint rather than an implementation detail.

Timing is part of cost because roadmaps shorten the useful life of a purchase. A technically strong system bought today may still be the wrong deal if the contract treats upgrades as vague aspirations and acceptance metrics as soft language. Buyers should therefore buy present capability but negotiate the future path. Upgrade clauses, refresh options, and buyer-defined acceptance tests are often worth more than a modest discount on the purchase price \citep{IBMRoadmap2026,IonQRoadmap2025,QuantinuumRoadmap2024,PasqalRoadmap2025,QuandelaRoadmap2026,DWaveGateRoadmap2026}.

\section{Questions that decide whether a purchase ages well}

Good procurement converts marketing language into contract language. Table~\ref{tab:questions} collects the questions that do that work. Each question tests whether the institution is buying a present capability or a future promise. Each question also maps to one of the failure modes highlighted earlier by Figure~\ref{fig:workflow}, Figure~\ref{fig:platforms}, and Figure~\ref{fig:cost}. The purpose is not to create paperwork. It is to force the institution and vendor to define what success means before the system becomes hard to change.

The first question is always what is delivered now. Buyers should write delivered capability, access conditions, uptime commitments, benchmark procedures, and support scope into the agreement and keep roadmap language outside acceptance criteria. The second question is what the site must support. On-premises pages from IQM, QuEra, D-Wave, AQT, Atom, OQC, Quandela, and ORCA all show, in different ways, that site requirements can determine feasibility long before the first workload runs \citep{IQMSpark2023,QuEraOnPrem2026,DWaveSystems2026,AQTSystems2025,AtomTechnology2026,OQCDeployments2026,QuandelaMosaiQ2026,ORCATechnology2026}. A system that fits the budget but not the building is not an available option.

Institutions should also ask what happens after the first year. How long is software support guaranteed? What API changes are planned? What training is included for new users rather than only the initial team? What is the policy for spare parts, preventive maintenance, and end-of-life transition? These questions rarely appear in vendor headlines, but they determine whether a purchase remains usable as staff, students, and workloads change.

The hidden question underneath Table~\ref{tab:questions} is blunt: who owns the hard parts after delivery? If the answer is not clear for calibration, maintenance, software updates, cybersecurity, user onboarding, documentation, and upgrade rights, then the institution does not yet know what it is buying. In quantum procurement, operational ambiguity is a form of technical risk.

\section{Discussion}

The procurement problem is simpler than the platform war. Table~\ref{tab:layers} and Figure~\ref{fig:workflow} show that the real first choice is the capability layer. Institutions that choose the layer first usually make better decisions than institutions that shop by qubit count, prestige, or roadmap headlines. Capability layers map directly onto budgets, users, and support obligations. Vendor headlines do not.

The market does not support one universal buying rule. Figure~\ref{fig:platforms} and Table~\ref{tab:roadmaps} show why. Superconducting systems lead in ecosystem breadth and access maturity. Trapped ions lead in premium logical-depth positioning. Neutral atoms lead in fast-moving scaling and error-correction narratives. Annealing remains the clearest fit-for-optimization tool. Photonics divides between general-purpose ambition and application-shaped deployment. The right answer depends on mission, not on slogans.

A useful way to read the market is by institutional starting point. For most teaching-focused institutions, the default is multi-vendor cloud access supplemented, where useful, by instructional hardware. For research universities with active quantum groups, a modest local instrument becomes rational when local calibration, methods development, or graduate training are core outcomes. For national laboratories and secure HPC centers, strategic local ownership becomes plausible when sovereignty, low-latency classical coupling, or protected data handling dominate the mission. For industrial R\&D groups, reserved access tied to a benchmark suite is often more rational than immediate ownership. These are starting presumptions, not rigid rules, but they are a better guide than qubit count alone.

A staged entry path is usually the most rational path. Table~\ref{tab:limited} shows that institutions can build a serious program by moving from access to teaching and then to local instrumentation in sequence. This is not hesitation. It is a method for buying evidence before buying infrastructure. In a market with rapid technical change and uneven support models, staged entry is often the most professional form of risk management.

Long-term value depends on what happens after delivery. Figure~\ref{fig:cost} and Table~\ref{tab:questions} make that point from two directions. The cost model shows that staffing, service, portability, and upgrade reserve matter as much as acquisition. The due-diligence table shows how those recurring obligations should appear in the contract. A machine becomes a platform only when the institution can operate it, support it, and keep it relevant.

This manuscript relies on public scientific literature and public commercial disclosures. That source base makes the comparison transparent, but it also leaves gaps where customer-specific pricing, service-level commitments, and negotiated support terms are not public. Buyers should therefore use this framework as a decision aid, then validate final numbers, benchmark procedures, and acceptance conditions during procurement.

\section{Conclusion}

Buying a quantum computer is not mainly a hardware choice. It is a decision about what level of quantum capability an institution can use, support, and sustain. For most buyers, the right path starts smaller and moves in stages. The main procurement mistake is to rank platforms before defining the mission. Buyers should first identify the target workloads, users, security needs, support model, site readiness, and software environment. Only then does platform choice become meaningful. Roadmaps are useful signals, but they are not evidence of delivered capability. Vendors should be judged first by what users can access now, how well the system fits institutional workflows, and what support and upgrade terms are contractually clear. Cost must also be read as total cost of ownership, not acquisition price alone. Staffing, maintenance, training, software support, facilities, and access reliability can matter as much as the machine itself. The platform comparison leads to no universal winner. Superconducting systems offer the broadest ecosystem, trapped ions remain strong for high-fidelity and logical-depth research, neutral atoms are advancing rapidly, annealing is credible for optimization-centered use, and photonics spans both general-purpose and application-driven paths. The right choice depends on mission fit. The practical answer is therefore simple. Institutions should buy the smallest layer of quantum capability that meets a real objective, builds useful expertise, and preserves flexibility. In most cases, staged entry is a stronger strategy than premature scale.

\section*{Notes}
The authors declare no conflicts of interest.

\bibliographystyle{unsrtnat}
\bibliography{refs}

\end{document}